\documentclass[nonacm,acmsmall,screen,letterpaper]{acmart}
\usepackage{soul}
\usepackage{subcaption}
\usepackage{tabularx}
\usepackage{xspace}

\usepackage{tikz}
\usetikzlibrary{positioning}
\usepackage{csquotes}

\usepackage{MnSymbol,bbding,pifont}
\usepackage{wrapfig} 
\usepackage{blindtext} 

\AtBeginDocument{%
  \providecommand\BibTeX{{%
    \normalfont B\kern-0.5em{\scshape i\kern-0.25em b}\kern-0.8em\TeX}}}

\setcopyright{acmcopyright}
\copyrightyear{2023}
\acmYear{2023}

\begin{document}

\title[Sub-Standards and Mal-Practices]{Sub-Standards and Mal-Practices: Misinformation's Role in Insular, Polarized, and Toxic Interactions on Reddit}

\author{Hans W. A. Hanley}
\email{hhanley@cs.stanford.edu}
\affiliation{%
  \institution{Stanford University}
  \streetaddress{450 Serra Mall}
  \city{Stanford}
  \state{California}
  \country{USA}
  \postcode{94305}
}


\author{Zakir Durumeric }
\email{zakir@cs.stanford.edu}
\affiliation{%
  \institution{Stanford University}
  \streetaddress{450 Serra Mall}
  \city{Stanford}
  \state{California}
  \country{USA}
  \postcode{94305}
}


\begin{abstract}
In this work, we examine the influence of unreliable information on political incivility and toxicity on the social media platform Reddit. We show that comments on articles from unreliable news websites are posted more often in right-leaning subreddits and that within individual subreddits, comments, on average, are 32\% more likely to be toxic compared to comments on reliable news articles. Using a regression model, we show that these results hold after accounting for partisanship and baseline toxicity rates within individual subreddits.  Utilizing a zero-inflated negative binomial regression, we further show that as the toxicity of subreddits increases, users are more likely to comment on posts from known unreliable websites. Finally, modeling user interactions with an exponential random graph model, we show that when reacting to a Reddit submission that links to a website known for spreading unreliable information, users are more likely to be toxic to users of different political beliefs. Our results collectively illustrate that low-quality/unreliable information not only predicts increased toxicity but also polarizing interactions between users of different political orientations. 

\end{abstract}



\maketitle

\section{Introduction} 
\vspace{1pt}
\noindent\fbox{%
    \parbox{.99\columnwidth}{%
        \textbf{Content Warning}: This paper studies online toxicity. When
        necessary for clarity, this paper quotes user content
        that contains profane, politically inflammatory, and hateful content.
    }%
}
\vspace{1pt}\\
Over the last decade, misinformation, incivility, and political polarization have corroded the public's trust in democratic institutions~\cite{chen2022misleading, gaughan2016illiberal, gervais2015incivility, borah2013interactions, goovaerts2020uncivil}. 
Despite their shared roles in disrupting discourse and stoking political division, misinformation, online toxicity, and polarization are separate phenomena, and their complex interaction remains debated and somewhat unclear~\cite{tucker2018social,mathew2020hate,goldenberg2020digital,shen2020viral,cacciatore2016end,darmstadt2019murder,cinelli2021dynamics,dori2021restoring,vicario2019polarization}. For instance, recent work from Quattrociocchi et~al.~\cite{quattrociocchi2022reliability} found that on that X (formerly Twitter), toxic language is equally distributed across conversations regardless of the presence of reliable or unreliable news. Similarly, Cinelli et~al.\ found that  ``there are no significant differences between the proportions of hate speech detected in comments on videos from questionable and reliable channels'' on YouTube~\cite{cinelli2021dynamics}. In contrast, Mosleh et al.~\cite{mosleh2024misinformation,mosleh2022measuring} found that false headlines on Twitter are correlated with increased toxicity and Dicicco et~al.~\cite{dicicco2020toxicity} found that throughout the COVID-19 pandemic, conspiracy theories emerged amongst users who regularly employed toxic language.

In this work, we investigate the interplay of toxicity, partisanship, and unreliable information in a more controlled environment: Reddit. In contrast to prior work, which has studied unstructured platforms like Twitter and YouTube~\cite{park2015comparing,rotman2009community}, Reddit communities have relatively distinct and stable political and toxicity norms~\cite{rajadesingan2020quick,kumar2023understanding,mamakos2023social}, allowing for more direct study of the complex interplay of toxicity, partisanship, and unreliable information. By investigating individual communities, quantifying their level of political engagement, and identifying internal differences between and within them, we analyze the extent to which partisanship, polarization, and unreliable news predict increased toxicity. Furthermore, in contrast to prior work, which has been limited to explicitly political settings, we analyze a diverse set of subreddits, measuring the influence of misinformation on toxicity while accounting for the ``politicalness'' of each community~\cite{mamakos2023social}.
Concretely, we ask the following research questions:

\begin{enumerate}
    \item \textit{Do Reddit posts linking to articles from unreliable websites have increased toxicity in their engagement? How do subreddit norms (\textit{e.g.}, political partisanship) predict toxicity?}
    \item \textit{Does unreliable news exacerbate toxic interactions between users with political partisanship differences (\textit{i.e.,} affective polarization)?}
   
\end{enumerate}

\noindent
To answer these questions, we measure the levels of toxicity, political partisanship, and propensity to post articles from websites known to spread misinformation on Reddit over 18~months (January~2020 to June~2021). We determine the number of toxic comments within each subreddit and from individual users using the Google Jigsaw API~\cite{perspectiveapi}, a commonly deployed classifier for identifying toxic language. Then, utilizing a Word2Vec approach from Waller {et~al.}~\cite{waller2021quantifying}, we approximate the partisanship and ``politicalness'' of a subset of subreddits and users along the US left--right political spectrum. Finally, we utilize previously curated lists of reliable and unreliable news sites to determine the levels at which communities and users link to websites known to spread misinformation. From these calculations, we analyze the relationships between toxicity, political partisanship, and misinformation:

\vspace{2pt}\noindent
\noindent
\textbf{RQ1: Toxicity, Partisanship, and Unreliable Information.} 
We first determine whether there are distinct levels of user political partisanship and toxicity in the comments that respond to articles from unreliable versus reliable news outlets.  We find that comments posted on articles from unreliable websites are on average 32\% more toxic within individual subreddits and 25\% more toxic across Reddit as a whole than comments responding to reliable websites. 
Fitting a linear regression against the average toxicity of users' comments, we find that the ``politicalness''/level of political engagement, each subreddit/community's toxicity norms, and prominently whether a post involves a low-reliability news website predict the toxicity of conversations. Finally, we show that as subreddits become more toxic, users are more likely to comment on unreliable news articles. In contrast, submissions linked to reliable sources are less likely to be engaged with in more toxic communities.


\vspace{2pt}\noindent
\noindent
\textbf{RQ2: Engagement with Unreliable News Source's Predicting Inter-Political Strife.} Having identified that users who comment on unreliable sources are more likely to post toxic comments than those who respond to reliable website posts, we examine the role of political partisanship in these toxic interactions. We find that users who comment under Reddit submissions to unreliable sources have a higher rate of inter-partisan toxicity compared to users who comment under reliable sources (1.38 odds ratio) and on Reddit generally (1.19). Indeed, users who comment on unreliable domain submissions are more likely to respond to users of different political views in a toxic manner and to reciprocate toxic comments aimed at them.

\vspace{3pt}\noindent
Altogether, we show unreliable websites' role in promoting toxicity on Reddit. Our work, one of the first to examine the relationship between unreliable news sources, toxicity, and political partisanship within and between different communities of varying levels of political engagement illustrates the need to fully understand the complex interactions between these phenomena so that platforms can better understand and address toxicity online.

\section{Background \& Related Work}

In this section, we detail key definitions, provide background on Reddit, and overview prior works that analyze the effects of misinformation, toxicity, and political polarization on social media.

\subsection{Terminology}\label{sec:misinformation-defintion}

Building on extensive prior work on misinformation, toxicity, and political polarization~\cite{hanley2022no,thomas2021sok,gruzd2020going,cinelli2020echo}, we utilize community-accepted definitions of the following terms:

\vspace{2pt}
\noindent
\textbf{Reliable and Unreliable Domains.} As in previous studies~\cite{jiang2018linguistic,guess2018selective,huang2015connected,lewandowsky2012misinformation,weeks2015emotions,hanley2022no,allcott2019trends}, we define \textit{misinformation} as information that is false or inaccurate regardless of author intention. Similarly, we define \textit{unreliable domains} as websites that regularly publish false information about current events and that do not engage in journalistic norms such as attributing authors and correcting errors~\cite{hanley2022no,zannettou2017web,hounsel2020identifying,sharma2022construction,abdali2021identifying,paraschiv2022unified,cheng2021causal,allcott2019trends}. Conversely, we define \textit{reliable domains} as websites that generally adhere to journalistic norms including attributing authors and correcting errors; altogether publishing mostly true information~\cite{hanley2022no,zannettou2017web,hounsel2020identifying}.



\vspace{2pt}
\noindent
\textbf{Online Toxicity and Incivility.} Given our use of the Google Jigsaw Perspective API~\cite{perspectiveapi}, we use their definition of toxicity:  ``\textit{(explicit) rudeness, disrespect or unreasonableness of a comment that is likely to make one leave the
discussion.}'' 

\vspace{2pt}
\noindent
\textbf{Partisanship.} We define partisanship as users' and communities' place on the US left--right political spectrum~\cite{robertson2018auditing}. We note the limitation of this definition given the variety of political views within the US. However, in line with previous work~\cite{saveski2022perspective,saveski2022engaging,hanley2022golden}, we utilize this definition, which largely fits much of US-centered political discussion, to understand how right-leaning and left-leaning users and communities interact with one another and news.

\vspace{2pt}
\noindent
\textbf{Affective Political Polarization:} 
Affective political polarization is the tendency of individuals to distrust and be negative to those of different political beliefs while being positive towards people of similar political views~\cite{druckman2021affective}. 

\subsection{Reddit}
Reddit is an online social media platform composed of millions of subcommunities known as subreddits~\cite{RedditMetrics,chandrasekharan2018internet}. Subreddits are dedicated to specific topics, ranging from politics (r/politics) and science (r/science) to Pokemon (r/pokemon). Depending on the community, users can submit news articles, opinions, images, and memes as \textit{submissions}. Underneath these submissions, other users can comment or reply to comments from other users. Anyone can create a subreddit and subreddits are moderated by Reddit content policies, subreddit-specific rules, and implicit community norms~\cite{chandrasekharan2018internet,fiesler2018reddit,jhaver2019did}. Subreddit norms vary widely~\cite{weld2022makes} and encompass political behaviors, tolerance to misinformation, and toxic behavior~\cite{rajadesingan2020quick,chandrasekharan2018internet,weld2022makes,jhaver2019did}. 



\subsection{Partisanship and Polarization}
People, both in real life and on the Internet, tend to associate with like-minded people~\cite{kamin2019social,huckfeldt1995political,halberstam2016homophily,barbera2014social,barbera2015tweeting,quattrociocchi2011opinions,hanley2022special}. Wojcieszak {et~al.}~\cite{wojcieszak2009online} find that while the majority of political discussions online are between participants who share the same viewpoint, many users \textit{do} enjoy conversations with people with different viewpoints~\cite{stromer2003diversity}. 
Despite this, past works have found that social media platforms are one of the drivers of political polarization ~\cite{cacciatore2016end,kamin2019social,heltzel2020polarization,pew2017partisan}. Sunstein, Garett {et~al.}, and Quattrociocchi {et~al.} all argue that the ``individualized'' experience offered by social media platforms comes with the risk of creating ``information cocoons'' and ``echo chambers'' that accelerate polarization~\cite{sunstein2018social,garrett2009echo,quattrociocchi2016echo}. Conover {et~al.}~\cite{conover2011predicting} find that 
Twitter/X's structure fosters increased levels of politically polarized conversations. Bessi {et~al.}~\cite{bessi2016users}, examining the behaviors of 12~million users, find that partisan echo chambers are driven by the algorithms of both Facebook and YouTube. Torres {et~al.}~\cite{torres2022manufacture} find the specific Twitter behavior of ``follow trains'' induce highly politically polarized behavior on the platform.

In a similar vein, prior work has found that the increased political polarization engendered by social media causes several negative downstream effects including the increased sharing of misinformation and toxic online behaviors. Imhoff {et~al.}~\cite{imhoff2022conspiracy}, for example, find that political polarization is associated with beliefs in conspiracy theories. Ebling {et~al.}~\cite{ebeling2022analysis} similarly find that political partisanship levels on social media are associated with medical misinformation about COVID-19. Other studies have further interrogated the adverse effects that social media has had on the democratic process due to the increased political polarization associated with social media~\cite{tucker2017liberation,tucker2018social,persily20172016,gron2020party}. 



\subsection{Misinformation}
Misinformation has increasingly become a major aspect of the conversations on social media~\cite{allcott2019trends,ghosh2018digital,funke2018fact}. Even after controlling for cascade size, Juul and Ugander find that false information spreads deeper and wider on Twitter/X than true information~\cite{juul2021comparing}. Furthermore, misinformation often convinces those who are exposed to it. A large percentage of US adults were exposed to misinformation stories by social media during the 2016 election~\cite{allcott2019trends} and many believed these false stories~\cite{allcott2017social,guess2018selective}. As COVID-19 spread throughout the world, online misinformation and conspiracy theories became a major hurdle to curbing its spread~\cite{romer2020conspiracy,sharma2022covid}.

To prevent the spread of misinformation, recent research has focused on tracking and stemming its flow~\cite{hanley2022no,tucker2018social}. For example, Mahl {et~al.}~\cite{mahl2021nasa}, track the spread of 10~conspiracy theories on Twitter, identifying one of the largest conspiracy theorist networks. Ahmed {et~al.}~\cite{ahmed2020covid} use a similar approach to track the spread of COVID-19 and 5G conspiracy theories. They find well-known misinformation websites were some of the largest sources spreading these conspiracy theories on Twitter. Gruzd~\cite{gruzd2020going} found that a single Tweet about how COVID-19 was a hoax, spurred an entire conspiracy theory, eventually prompting large groups of people to film their local hospitals to prove that COVID-19 was not real. In addition to network-based approaches, others have used advancements in natural language processing to identify and track misinformation. Hanley {et~al.}~\cite{hanley2022happenstance}, for example, utilize semantic search to identify and track Russian state-media narratives on Reddit. Fong {et~al.}~\cite{fong2021language} utilized linguistic and social features to understand the psychology of Twitter users that engaged with known conspiracy theorists. Finally, several works have performed in-depth case studies on the spread of specific false narratives: Wilson and Starbird {et~al.} look at the Syrian White Helmets on Twitter and Bär {et~al.} look at the spread of QAnon on Parler~\cite{wilson2020cross,https://doi.org/10.48550/arxiv.2205.08834}.   


\subsection{Toxicity}
Online toxicity takes many forms including threats, sexual harassment, doxing, coordinated bullying, and political incivility~\cite{freed2017digital,freed2018stalker,lima2018inside,thomas2021sok}.  Toxic comments, in particular,  are one of the most common forms of hate and harassment online~\cite{thomas2021sok} and 
are seemingly an inescapable part of social media~\cite{thomas2021sok,cuomo2019gender,kumar2021designing,nobata2016abusive,wulczyn2017ex}. Past studies have found that 41\% of Americans and 40\% of those globally have experienced bullying or harassment online~\cite{thomas2021sok, Duggan2017}. Facebook estimates that 0.14--0.15\% of all views on their platform are of toxic comments~\cite{facebook2022}. This type of incivility, in addition to damaging online conversations, has been found to also damage civil institutions~\cite{tucker2017liberation,borah2013interactions} having dangerous real-world implications. For example, Fink {et~al.}~\cite{fink2018dangerous} find that politically charged anti-Muslim hate speech on Facebook in Myanmar was a prominent aspect preceding the Rohingya genocide.

To limit toxicity, platforms have designed and implemented a variety of safeguards~\cite{perspectiveapi, TwitterEnforcement, facebook2022}. Other researchers have further performed in-depth studies on users' behavior to understand abusers and victims of abuse. For instance, Founta {et~al.}~\cite{founta2018large} identify a set of network and account characteristics of abusive accounts on Twitter. Hua {et~al.}~\cite{hua2020characterizing} look at properties of the accounts that have heavily negative interactions with political candidates on Twitter. Finally, Chang {et~al.}, Xia {et~al.}, Zhang {et~al.}, and Lambert {et~al.} all look at the set of causes that make conversations unhealthy or  toxic~\cite{zhang2018conversations,zhang2020quantifying,lambert2022conversational,xia2020exploring}. 

\subsection{The Interplay of Misinformation, Online Toxicity, and Political Polarization}
Several works have attempted to understand how political partisanship, online toxicity, and misinformation interact. Online toxicity, for instance, has been heavily associated with increased political polarization and misinformation~\cite{cinelli2021dynamics,tucker2018social}. Rajadesingan {et~al.}~\cite{rajadesingan2021political}, find that political discussions in non-overtly political subreddits often lead to less toxic conversations. Cinelli {et~al.}~\cite{cinelli2021dynamics}, show that misinformation about COVID-19 on YouTube promoted hate and toxicity. Chen {et~al.}~\cite{chen2022misleading}, utilizing network-based analysis, find that misleading online videos often lead to increased incivility in their comments. Separately, Rains {et~al.}~\cite{rains2017incivility} find that political extremism is a major factor in toxicity online. De Francisci Morales {et~al.}~\cite{de2021no} find, most markedly that the interaction of individuals of different political orientations increased negative conversational outcomes. Similarly, Kim {et~al.}, Kwon {et~al.}, and Shen {et~al.}  find that exposure to negative conversations increases observers' tendency to further engage in incivility~\cite{kim2019incivility,kwon2017offensive,shen2020viral}. Finally, Imhoff {et~al.}~\cite{imhoff2022conspiracy} find that political polarization is a key aspect of people's belief in false narratives. However, despite this panoply of research, it is unclear how political partisanship and toxicity interact in the presence of misinformation and across political environments. In this work, we seek to understand this dynamic. 

\subsection{Present Work}
While several previous works have studied partisanship and affective polarization~\cite{mamakos2023social,de2021no,efstratiou2022non}, finding evidence of inter-partisan hostility, these works has been limited to explicitly politically-oriented spaces and do not study the influence of unreliable information or misinformation. As shown by Rajadesingan et~al.~\cite{rajadesingan2021political} and Mamakos et~al.~\cite{mamakos2023social}, political discussions frequently take place in non-overtly political subreddits. Limiting the study of how partisanship and unreliable information affect users' discussions to only overly political subreddits, as in past works, can thus give an incomplete picture of user behavior. As found by Efstratiou et~al., different subreddits can have different ``echo chamber-like'' behaviors and inter-partisan discussions depending on their politicalness~\cite{efstratiou2022non}. 

Our work seeks to understand how partisanship and unreliable news sources that spread largely non-factual information contribute to this toxicity and user engagement in both political and non-political contexts and within individual subreddits/communities.  Given that our work quantifies the politicalness and other characteristics of a subreddit or a user utilizing the methodology outlined by Waller et~al.~\cite{waller2021quantifying}, we can account for this factor in contributing to toxicity and explore how unreliable sources interact in different subreddit environments and across different community standards. By examining how these unreliable and reliable sources differ in toxicity both within and between individual subreddits and across subreddits of different types of politicalness, we seek to understand the extent to which unreliable news promotes toxicity and engagement among users of different political orientations. 

\section{Datasets and Methods}\label{sec:methods}


In this section, we provide an overview of our datasets and describe how we calculate the political partisanship of users and subreddits, how we determine the toxicity of posts and comments, and how we identify user interactions with unreliable and reliable website sources. 

 \subsection{Reddit Dataset}\label{sec:reddit-methods}

We study 18~months of Reddit comments and submissions from January 2020 to June 2021, which we collect using Pushshift~\cite{baumgartner2020pushshift}. Altogether, we gather 2.2~billion comments and 491~million submissions. Each comment and submission includes its timestamp, author's username, subreddit, and the conversation thread where the comment was posted. We note that all data was collected before Pushshift fell outside Reddit's Terms of Service in April 2023. Using this data, we reconstruct the conversation threads for each user and subreddit. 


As in Kumar et~al.~\cite{kumar2023understanding}, given that many Reddit comments labeled as toxic are simply sexually explicit and contained within 18+ communities, we exclude 18+ subreddits from our study. As argued by Kumar et~al.~\cite{kumar2023understanding}, while toxic behaviors do occur within these subreddits, the explicit allowance of sexually explicit language leads to a large number of false positives, complicating analysis. In addition to filtering out 18+ subreddits, we limit our analysis to English-language misinformation and thus filter our dataset using the \texttt{whatlanggo} Go language library\footnote{\url{ https://github.com/abadojack/whatlanggo}} to only English-language comments. Finally, given the model that we utilize to detect toxicity, we limit our analysis to comments that are 15--300 characters in length~\cite{kumar2021designing}. Finally, to ensure that the user and subreddit characteristics that we extract are robust, we only calculate statistics for subreddits with at least 100~comments and users that posted at least 5 comments. Altogether, our final dataset consists of 327M~Reddit submissions, 1.6B~comments, and 15.5M~users from 57.2K~subreddits.

\subsection{Unreliable and Reliable Domain Dataset}

To analyze how users interact with misinformation, we first gather a set of unreliable and reliable websites (as a control). Specifically, we aggregate a list of unreliable/misinformation and reliable/authentic-news domains from Media-Bias/Fact-Check.\footnote{\url{https://mediabiasfactcheck.com/}} We consider websites as “unreliable” if their factfulness rating from Media-Bias/Fact-Check is ``Low'' or ``Very Low''; conversely, we consider a website as ``reliable'' if its factuality rating from Media-Bias/Fact-Check is ``Mostly Factual'', ``High'', or ``Very High''. We include ``Mostly Factual'' in this category given that it includes websites like cnn.com and washtingtonpost.com. To ensure consistency, we further cross-reference these two lists of websites against news websites previously gathered by Iffy News,\footnote{\url{https://iffy.news/index}} OpenSources,\footnote{\url{https://github.com/several27/FakeNewsCorpus}} Politifact,\footnote{\url{https://www.politifact.com/article/2017/apr/20/politifacts-guide-fake-news-websites-and-what-they/}} Snopes,\footnote{\url{https://github.com/Aloisius/fake-news}}  Melissa Zimdars,\footnote{\url{https://library.athenstech.edu/fake}} and Hanley et~al.~\cite{hanley2022golden}. Our final list of misinformation outlets consists of 1,054 websites, which encompass sites like theconservativetreehouse.com and infowars.com~\cite{hanley2022golden}.  Separately, our list of reliable news sites consists of 3,754~websites from across the political spectrum, including sites like cnn.com and nytimes.com.

\subsection{Approximating the Partisanship of Subreddits and Users}
To approximate the political partisanship of subreddits and Reddit users, we adopt the neural embedding approach described by Waller et~al.~\cite{waller2021quantifying,waller2019generalists}, which learns subreddit and user embeddings/vectors based on the interaction data of users within subreddits. This is such that a high cosine similarity between two users would indicate that the two users comment/post in similar or the same subreddits; conversely, a high similarity between two subreddits would indicate that they share similar user bases. By computing subreddit and user similarity scores along a political partisanship dimension created when training the Wor2Vec model, as in Waller et al.~\cite{waller2021quantifying}, this approach enables the approximation of the partisanship of users and subreddits. We utilize this approach as it allows us to avoid biases in previous manual labels of the political orientation of subreddits and because it allows us to label the orientation.  Specifically, as in  Waller et~al.~\cite{waller2021quantifying}, we apply the Word2Vec algorithm to our Reddit data where subreddits are treated as ``words'' and users are treated as ``contexts''. In this approach, every individual instance of a Reddit user commenting or submitting in a given subreddit is considered a word-context pair. Upon aggregating these word-context pairs, we subsequently train a Word2Vec using skip-gram with negative sampling outputting the vector embedding for each subreddit and for each user.  

From our vector embeddings, as specified by Waller et al.~\cite{waller2021quantifying}, we identify the political partisanship dimension elicited by the Word2Vec to then categorize the political orientation of individual subreddits and users. More concretely, after extracting our embeddings, we identify two similar communities that differ primarily in the our dimension of interest; in this case, r/democrats and r/conservative. From the Word2Vec embeddings $\mathbf{sr}_{r/democrats}$ and $\mathbf{sr}_{r/conservative}$ that we elicited from these subreddits, we then compute the political partisanship dimensional vector $\mathbf{pr}_{1} $ = $\mathbf{sr}_{r/democrats} - \mathbf{sr}_{r/conservative}$. To ensure that the political dimension that we are studying is not overly specific to our seed communities of  $\mathbf{sr}_{r/democrats}$ and $\mathbf{sr}_{r/conservative}$, we subsequently identify other pairs of similar communities whose difference vector has a high cosine similarity to our political partisanship dimensional vector $\mathbf{pr}_{1}$  (\textit{i.e.}, other pairs of communities that differ primarily in political partisanship direction). For example, in our work, other pairs of communities that differed primarily along our political dimension included: r/liberalgunowners and r/gunpolitics, r/climatechange and r/climateskeptics, and  r/askaliberal and r/askaconservative. As in Waller et~al.~\cite{waller2021quantifying}, we average thee vectors to get our final partisanship dimensional vector $\mathbf{pr}_{1} $ = $\mathbf{sr}_{r/democrats} - \mathbf{sr}_{r/conservative}$ using 10 unique political pairs. 

\begin{equation}
    \mathbf{pr} = \frac{1}{10}\sum_i^{10}\mathbf{pr}_{i}
\end{equation}

To project individual subreddits onto the political partisanship dimension, we compute the cosine similarity between a given community's Word2Vec embedding $\mathbf{sr}_{r/any\_subreddit}$ and the computed political partisanship dimension $\mathbf{pr}$ vector. To make these values more interpretable, as in Waller et~al.~\cite{waller2021quantifying}, we determine the z-scores for each community's projected value on the political partisanship dimension. This is such that a community with a z-score of -1 could be interpreted as having a leftward stance with a political partisanship level of 1 standard deviation below the mean subreddit. As in Waller et~al.~\cite{waller2021quantifying}, in addition to calculating the political partisanship of individual subreddits, by taking the sum of the vectors of our communities utilized to compute the political dimension, rather than the difference, we can also determine the ``political''-ness of individual subreddits and communities. This measure assesses the level of political engagement of a community or user, rather than pinpointing their position on the political spectrum. For example, the r/law subreddit, while not particularly partisan (-0.19$\sigma$), is over two standard deviations above the mean for politicalness (2.10$\sigma$).

\begin{figure}
\centering
\begin{minipage}[l]{0.49\textwidth}
\includegraphics[width=\columnwidth]{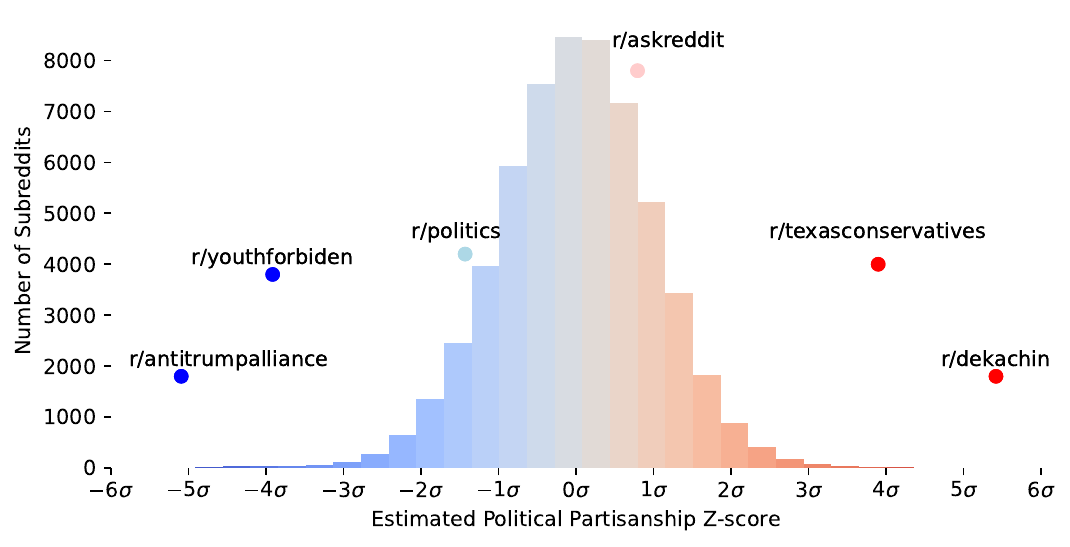}
\end{minipage}
\begin{minipage}[l]{0.49\textwidth}
\includegraphics[width=\columnwidth]{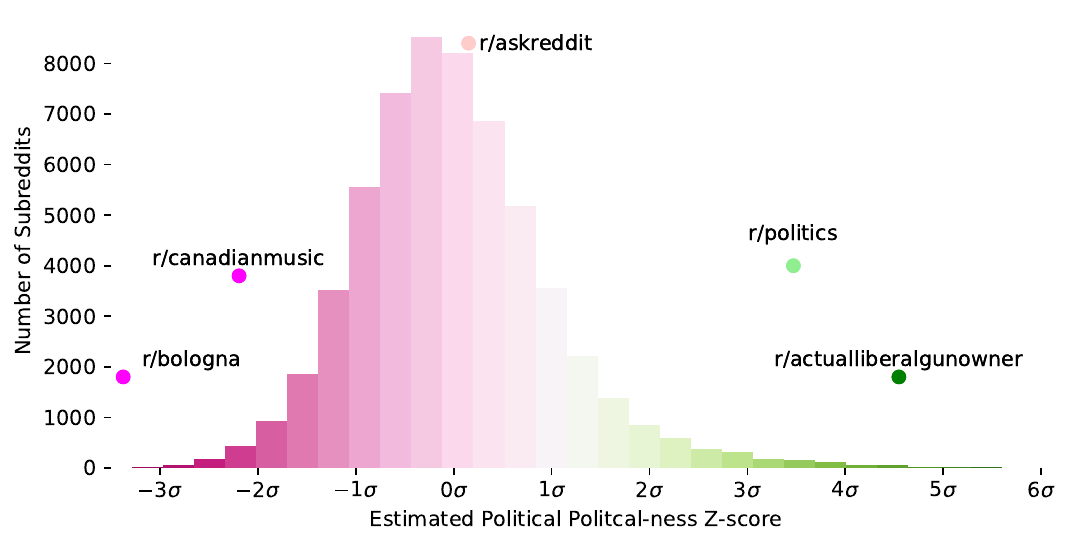}
\end{minipage}
\begin{minipage}[l]{1\textwidth}
\caption{{Subreddit political partisanship and politicalness distribution }--- We determine the political partisanship (where a subreddit falls on the US left/right political spectrum) and how political a subreddit is by utilizing Waller et~al.'s~\cite{waller2021quantifying} method for creating subreddit and user embeddings using an extension of Word2Vec~\cite{kumar2018community}. }
\label{figure:toxicities}
\end{minipage}
\end{figure}

We lastly note that given the many individual hyperparameters utilized within Word2Vec models (\textit{e.g.}, embedding size, down-sampling threshold, starting learning rate, \textit{etc...}), we perform a grid-search on these parameters and subsequently validate the political partisanship scores those of Waller et~al.~\cite{waller2021quantifying}. We select the model with partisanship scores that have the greatest Pearson correlation with those provided by  Waller et~al.\footnote{We do not utilize the political partisanship scores provided by Waller et~al.~\cite{waller2021quantifying} given that their study is limited to 10,006 ~ubreddits and given that they do not provide vectors or partisanship scores for individual users.} We detail the hyperparameters and the values that we optimize over in Appendix~\ref{sec:hyperparamters}.

\subsection{Identifying Toxic Comments and Approximating User and Subreddit Toxicity\label{sec:toxicity-classifction}} 

To approximate the toxicity of Reddit users and subreddits, we utilize the Perspective API, a set of out-of-box toxicity classifiers from Google Jigsaw~\cite{perspectiveapi} that has been utilized extensively in prior works~\cite{kumar2021designing,rajadesingan2020quick,saveski2021structure}. Each classifier takes comments as input and returns a toxicity score of 0.00--1.00; the closer a comment's score is to 1, the more likely the comment is to be toxic. In line with prior work, to consider a comment as toxic, we utilize a threshold of 0.80 on the \texttt{SEVERE\_TOXICITY} classifier~\cite{chong2022understanding,lambert2022conversational}. As found by Kumar~et~al.~\cite{kumar2021designing,kumar2023understanding}, utilizing this particular classifier, while limiting recall, provides an acceptable precision for identifying toxic online content.


\begin{figure}
\centering
\begin{minipage}[l]{0.49\textwidth}
\includegraphics[width=\columnwidth]{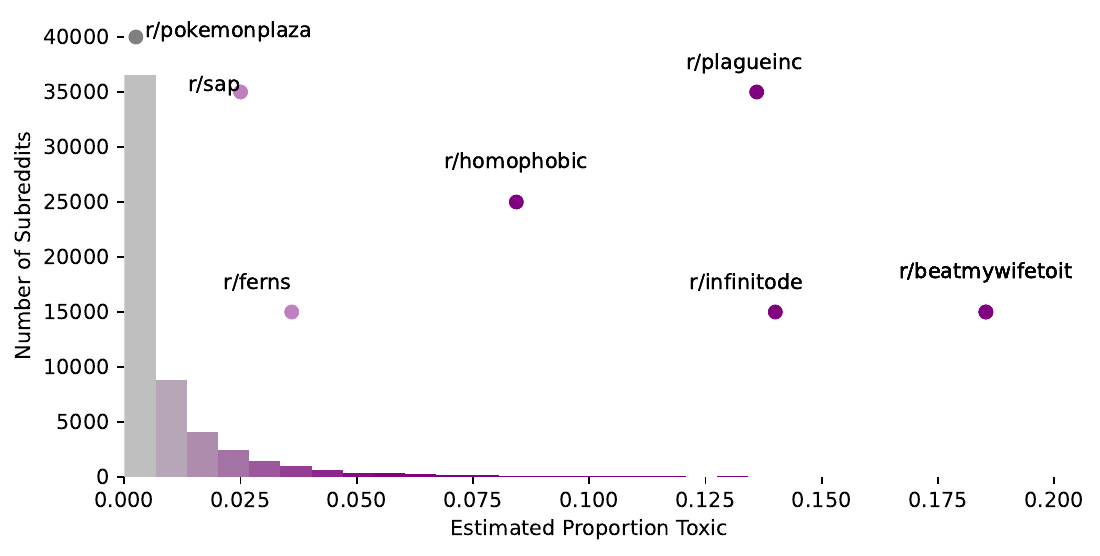}
\end{minipage}
\begin{minipage}[l]{0.49\textwidth}
\includegraphics[width=\columnwidth]{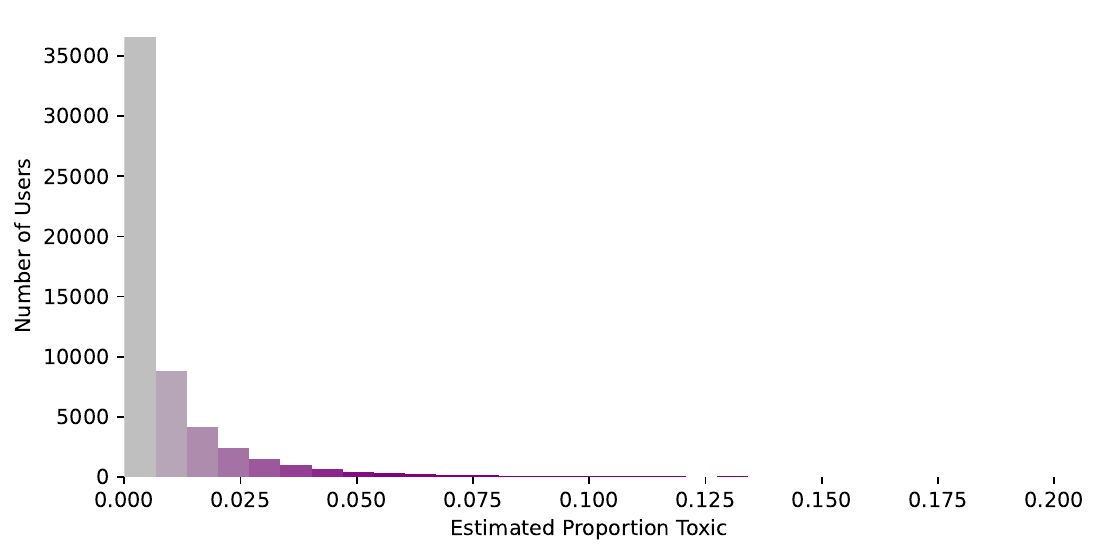}
\end{minipage}
\begin{minipage}[l]{1\textwidth}
\caption{{Subreddit and User Toxicity scores}---We determine the toxicity norms for subreddits with at least 100~comments and for users with at least 5~comments. Each user and subreddit has distinctive toxicity norms, posting toxic comments at different rates. At a threshold of 0.80, most users and the subreddit's usual comments/posts are not considered toxic by the Perspective API SEVERE\_TOXICITY classifier.}
\label{figure:toxicities}
\end{minipage}
\end{figure}

\subsection{Ethical Considerations} 
Within this work, we focus on identifying trends in how subreddits interact with misinformation, levels of toxicity, and levels of political polarization. While we do calculate toxicity and polarization levels for individual users, we do not analyze specific users, we do not publish their usernames, and we do not attempt to contact or deanonymize them. 
We note that the Reddit submissions and comments analyzed in this work were public and available through the Pushshift API~\cite{baumgartner2020pushshift}. 

\section{Toxicity and Partisanship in Misinformation Posts}\label{sec:rq1}
In this section, we examine the relationship between Reddit submissions utilizing unreliable information sources and their corresponding partisanship, toxicity, and user engagement (\textit{i.e.}, number of comments). Using reliable news submissions as a control and accounting for the types of subreddits where posts to unreliable sources appear, we measure whether Reddit posts that link to known unreliable information sources predict increased toxicity. After examining the distributional differences in several characteristics amongst the users and subreddits of unreliable and reliable news submissions, we finish this section by fitting a linear model and a negative binomial model to understand the degree to which each of these features predicts toxicity and user engagement on Reddit.

\begin{table}
    \tiny
    \centering
    \selectfont
    \begin{tabularx}{1.00\columnwidth}{lr|lr|lr|lr}
    \toprule
    Top Unreliable  websites &  \# Links & Top Reliable websites  &\# Links  & Top Unreliable Subreddits & \# Links  & Top Reliable Subreddits &\# Links  \\ \midrule
    oann.com & 188,678 &  nytimes.com& 493,032& r/TheNewsFeed & 133,600 & r/AutoNewspaper & 1,010,948 \\
    dailymailk.co.uk & 110,491 & cnn.com & 392,392& r/ConservativeNewsWeb & 64,565 & r/politics & 426,931\\
    rt.com & 27,347  & reuters.com& 245,633 & r/OneAmericaNews & 54,138& r/news & 208,612\\
    wnd.com & 25,732 & thehil.com &  219,826 & r/trendandstyle & 47,171& r/worldnews & 195,644 \\
    newsmax.com & 25,204 & cnbc.com & 179,157 & r/StateoftheUnionNONF & 27,232& r/Coronavirus & 178,555\\
    americanthinker.com &22,247   &nbcnews.com  & 174,430 & r/Conservative & 22,859 & r/nofeenews & 92,815\\
    sputniknews.com & 19,736 & yahoo.com& 164,489& r/StonkFeed & 16,941&r/nytimes & 89,795\\
    rumble.com & 17,172 & usatoday.com & 147,323 & r/TheBlogFeed & 15,543& r/NoFilterNews & 85,960\\
    zerohedge.com & 15,409 & washingtonpst.com & 128,579 & r/conspiracy & 13,510& r/NBCauto & 83,361\\
    bitchute.com & 12,788 &latimes.com & 124,742 & r/boogalorian & 8,730& r/CNNauto & 79,436\\
    \bottomrule
    \end{tabularx}
    \vspace{1pt}
  \caption{Top mainstream and websites hyperlinked within Reddit Submission and the top subreddits with unreliable websites and reliable websites hyperlinked. Altogether, within our set of studied 57K subreddits, we identify 633,585 submission hyperlinks to our set of unreliable news websites and a total of 7,546,917 submission hyperlinks to our set of reliable news.   }
   \vspace{-15pt}
   \label{table:top-websites-hyperlinked}
\end{table}

To understand the characteristics of users and communities that interact with unreliable sources, we identify submissions that link to our 1,057~unreliable and 3,754~reliable websites. Altogether, we find 633.59K~submissions of unreliable news websites and a corresponding set of 5.29~million comments and 7.55~million submissions that link to our set of reliable websites and 267~million corresponding comments. We list the most frequently linked websites and subreddits that most commonly link to our sets of sites in Table~\ref{table:top-websites-hyperlinked}. 
Altogether, hyperlinks to unreliable websites were posted in 9,462~subreddits and to reliable websites in 29,673~subreddits (8,611~subreddits had links to both). The difference in the magnitude of submission is likely due to the greater popularity and widespread appeal of reliable mainstream news compared with alternative, fringe websites~\cite{hanley2022golden}. Indeed, utilizing the Alexa Top Million list from March 1, 2021~\cite{amazon-top-mil}, we find that 991~reliable news websites  (26.39\%) were in the top 100K~websites compared to 139~unreliable websites (13.19\%).  

For the rest of this section, while using partisanship, politicalness, and toxicity averages computed from our full Reddit dataset (see Section~\ref{sec:methods}), we analyze the set of Reddit submissions and Reddit comments that involve unreliable and reliable website submissions. We additionally remove AutoModerator comments and comments from accounts labeled as ``bots.''

\subsection{Differences Between Unreliable and Reliable Website Submissions}\label{sec:toxicity}

Across our dataset, we find that 1.26\% of all comments within our datasets were classified as toxic (\textit{i.e.}, Perspective \texttt{SEV\_TOX} score $>$0.80), 1.24\% of comments under reliable website submissions were considered toxic, and 1.55\% of comments on unreliable submissions (a 25\% relative increase). However, as previously mentioned, these comments are largely posted in different communities on Reddit and likely by different users. Performing a comparison across individual subreddits, we find that there remains a mean absolute percentage increase of 0.35\% (32.2\% relative increase) in toxicity ($p< 1\times 10^{-16}$) for toxicity on unreliable news articles compared to reliable news articles.   In this section, we thus determine the differences between subreddits and users that interact with reliable versus unreliable news to understand this increase in toxicity. 



\begin{table}
    \scriptsize
    \centering
    \selectfont
\setlength{\tabcolsep}{3.5pt}
    \begin{tabularx}{0.59\columnwidth}{lrrr}
    \toprule
    &  \textbf{Unreliable} &   \textbf{Reliable} & \textbf{Cohen's D}  \\ \midrule

   Avg. Subreddit Partisanship &  0.96$\sigma$ & -0.17$\sigma$ &0.79   \\ 
   Avg. Subreddit Politcalness & 2.37$\sigma$ &  3.12$\sigma$  &-0.47  \\
   Avg. Subreddit Toxicity &  2.01\%   &   1.40\% & ---\\ \midrule
      Avg. Submitter Partisanship & -0.04$\sigma$ & -0.19$\sigma$ & 0.19  \\
   Avg. Submitter Politicalness & -0.01$\sigma$ & 0.49$\sigma$ & -1.42 \\ 
  Avg. Submitter Toxicity & 0.93\% &  0.90\% & ---  \\
    Avg. Submitter Account Age (Years) & 2.57 &  4.32 & -0.54\\\midrule
   Avg. Commenter Partisanship & 0.56$\sigma$ &0.09$\sigma$ & 0.57  \\
   Avg. Commenter Partisanship Var. & 0.45 & 0.48 & -0.07  \\ 
   Avg. Commenter Politicalness & 0.20$\sigma$ & 0.26$\sigma$ &-0.20 \\ 
    Avg. Commenter Politicalness Var. & 0.13 & 0.15 & -0.19  \\ 
    Avg. Commenter Toxicity & 1.48\% &  1.36\% & ----  \\
    Avg. Commenter Account Age (Years) & 4.88 & 5.25& -0.12 \\\midrule
    \% Removed Comments &   2.01\% &2.82\% &  ---\\ 
    \% Mod/Admin Involved &   16.74\% &16.26\% &  ---\\ 
   \bottomrule \\
    \end{tabularx}
  \caption{We determine different characteristics of the subreddits, commenters, and submitters that interact with reliable and unreliable website submissions and subsequently determine the Cohen's effect size between these values for unreliable news submissions and reliable news submissions. We perform Mann-Whitney U tests to ensure that the differences in the averages between unreliable and reliable website submissions are significant. We perform two-sample proportion tests for the percentages. Note, we performed a Bonferonni correction to assess whether values were significant, but all p-values tested were $p< 1\times 10^{-16}$ and significant. }
   \vspace{-15pt}
   \label{table:submitter-commeter-info}
\end{table}

\vspace{2pt}
\noindent
\textbf{Subreddits.} As seen in Table~\ref{table:submitter-commeter-info}, on average, the subreddits where unreliable website submissions are posted are 1.13~standard deviations more right-leaning on the US political spectrum than those of reliable websites. This accords with previous research that has found that right-leaning users and ecosystems are more likely to spread misinformation~\cite{jost2018ideological}. However, we also observe that unreliable website submissions tend to be posted in subreddits that are typically 0.75~standard deviations less political than reliable website submissions. For example, r/StreetFighter, a subreddit dedicated to the video game Street Fighter (politcalness=-0.47$\sigma$) contained 409~submissions to 4chan.org and r/MMA (politcalness=-0.43$\sigma$), had 81~links known Russian propaganda website rt.com~\cite{center2020pillars} and far-right conspiracy site infowars.com~\cite{van2020lizards}. Unreliable website submissions tend to be in subreddits with higher average toxicity (2.01\% vs.\ 1.40\% of comments), which may explain the higher likelihood of toxic comments in response to misinformation posts.

\begin{figure}
\begin{subfigure}{.49\textwidth}
  \centering
 \includegraphics[width=1\columnwidth]{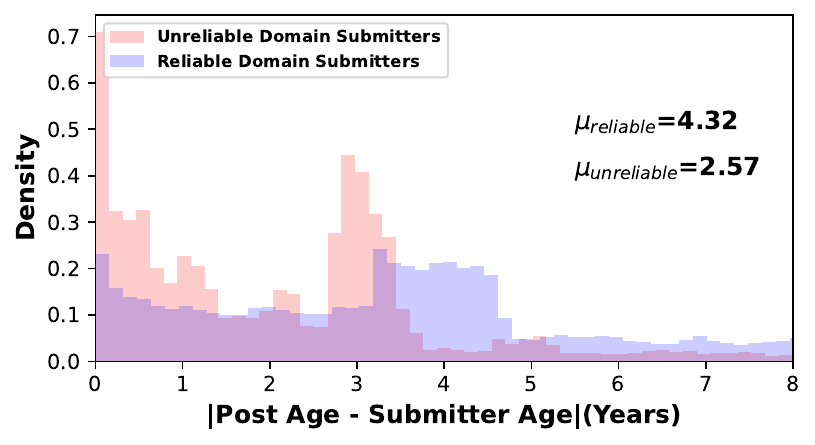} 
  \caption{Age of Submitters}\label{figure:creation_vs_toxicity}
\end{subfigure}
\begin{subfigure}{.49\textwidth}
  \centering
  \includegraphics[width=1\columnwidth]{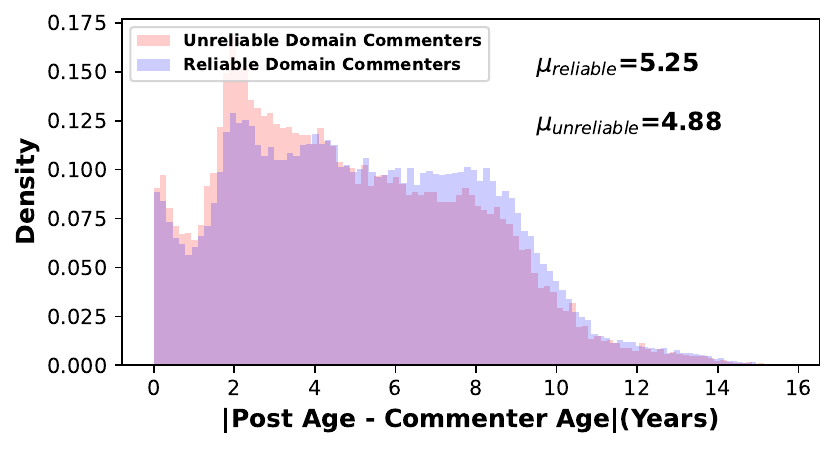}
    \caption{Age of Commenters\label{figure:age_of_commenters}}
\end{subfigure}
 \caption{Younger accounts are much more likely to submit and comment on unreliable website submissions.\label{fig:ages}}
 \vspace{-10pt}
\end{figure}

\vspace{2pt}
\noindent
\textbf{Submitters.}
In line with prior work, we find that users who submit unreliable websites articles as Reddit submissions tend to be more right-leaning (-0.04$\sigma$ vs.\ -0.19$\sigma$), tend to be less political (-0.01$\sigma$ vs.\ 0.49$\sigma$), tend to have slightly more toxic comments (0.93\% 0.90\%), and tend to have younger accounts (Table~\ref{table:submitter-commeter-info}). Performing a subreddit pairwise comparison (\textit{i.e.}, comparing the users who submitted unreliable websites in one subreddit to the users who also submitted reliable websites in the \emph{same} subreddit), we indeed find that users that submit unreliable websites tended to be more right-leaning (Cohen's D = 0.26, $p<1 \times 10^{-16}$ using the paired Wilcoxon signed-rank test), were very slightly more political  (Cohen's D = 0.01, $ p <1 \times 10^{-16}$), and were slightly more toxic overall  (0.12\% absolute percentage increase, $ p <1 \times 10^{-16}$).


We further observe that submitters of unreliable website hyperlinks tend to have younger accounts. As argued elsewhere, when posting inflammatory, revealing, or otherwise sensitive information~\cite{kumar2023understanding,ammari2019self,leavitt2015throwaway}, Reddit users often utilize disposable ``throw-away'' accounts that are used only to post this information anonymously. Indeed, as seen in Figure~\ref{fig:ages}, within our dataset,  we find that while only 0.88\% of reliable website submissions are submitted within the first week of an account's lifespan, 2.64\% are submitted in the first week for unreliable websites (we perform a proportion test and find this difference to be significant $p< 1\times 10^{-16}$).

\vspace{2pt}
\noindent
\textbf{Commenters.} Commenters on unreliable website submissions tend to be slightly more right-leaning (0.56$\sigma$ vs.\ 0.09$\sigma$), slightly less political (0.20$\sigma$ vs.\ 0.25$\sigma$), but slightly more toxic (1.48\% vs.\ 1.36\%). Performing a subreddit pairwise comparison (\textit{i.e.}, comparing the users that commented on unreliable websites in one subreddit to the users who commented on reliable websites in the \emph{same} subreddit), we find that the users who comment on unreliable websites have no significant difference (using the paired Wilcoxon signed-rank test) in partisanship nor toxicity, but do differ slightly in politicalness (Cohen's D = -0.07). We thus see that after accounting for the subreddit, \emph{that is largely the same type of users that comment on unreliable and reliable website submissions within a given subreddit}. Despite seeing that within subreddits the users of similar partisanship and toxicity post on unreliable and reliable news submissions, again performing this subreddit pairwise comparison, we find as previously reported that there is a mean absolute percentage increase of 0.35\% (32.2\% relative increase) in toxicity ($p< 1\times 10^{-16}$) for unreliable submissions within each subreddit. This illustrates that despite similar users participating in conversations surrounding unreliable and reliable news within a given subreddit, unreliable news comments tend to have more toxic language. 


As for submitters (Figure~\ref{fig:ages}), we find that commenters on unreliable website submissions have younger accounts than those for reliable website submissions (4.88 years vs.\ 5.25 years). We note that, as with submitters, this may partially explain the increased toxicity of unreliable submission commenters. Plotting the age of accounts versus the proportion of toxic comments in Figure~\ref{figure:creation_vs_toxicity}, we observe that age is indeed correlated with the toxicity of commenters, with (as expected) unreliable news websites having the highest toxicity overall regardless of the age of the account. 

\vspace{2pt}
\noindent
\textbf{Moderation and Removed Comments.} A potential confounder that can cloud our analysis is the activity of Reddit moderators. Reddit moderators are members of particular subreddit communities who help set rules and norms and help moderate content~\cite{almerekhi2020investigating}. When a moderator on the Reddit platform removes a comment, the comment is replaced with ``[removed]'' and other Reddit users can no longer view the comments. Altogether, 14,642 comments were removed from our set of unreliable website submissions and 3,305,138 comments were removed from our set of reliable website submissions. As seen in Table~\ref{table:submitter-commeter-info}, on average, reliable website submissions are more moderated compared to unreliable website submissions (with an average of 2.00\% comments being removed compared to 2.82\%). However, again performing a subreddit-wise pairwise comparison, we find that within the subreddits where both reliable and unreliable submissions appear, unreliable news commenters are actually moderated more heavily (Cohen's D = 0.37, $p< 1\times 10^{-16}$ using the paired Wilcoxon signed-rank test). This indicates, within subreddits that have both unreliable and reliable domain hyperlinks that unreliable ones are moderated more heavily; conversely, outside of these subreddits, these unreliable website submissions are moderated more leniently. For example, within the r/bicycling subreddit, while 2.01\% of reliable domain comments were removed, 19.35\% of unreliable domain comments were removed. In contrast within the r/bitchute, where there were no comments on reliable news domain hyperlinks, only 0.50\% of comments were removed (BitChute is an alternative to YouTube known for hosting toxic and conspiratorial content~\cite{trujillo2020bitchute}). 

 We lastly examine the cases where a moderator left a comment or interacted with users in the subreddit. As seen in Table~\ref{table:submitter-commeter-info}, across all our submissions, moderators were involved in slightly more submissions in unreliable domain submissions, either as the submitter or as a commenter. We find that the comments of submissions that had a moderator/admin involved were less toxic than those that did not (1.04\% vs. 1.66\%  for unreliable website submissions; 0.70\% vs 1.12\% for reliable website submissions). Performing a subreddit-wise pairwise comparison on the proportions of submissions per subreddit that had moderator involvement, we again see that unreliable news websites were very slightly more likely to have a moderator involved (Cohen's D =0.05,  $p< 1\times 10^{-16}$ using the paired Wilcoxon signed-rank test).  

\vspace{2pt}
\noindent
\textbf{Summary.}
In this section, we showed that links to websites known to spread unreliable information are correlated with higher toxicity: toxic comments under unreliable website submissions are posted at a rate of 1.55\% while toxic comments in response to reliable website submissions are posted at a rate of 1.24\%. Within individual subreddits, we find that on average, the toxicity rate increases on average by an absolute 0.35\% (32.2\% relative increase). In addition, we observed that users who post and comment on misinformation are right-leaning.

\subsection{Prediction of Toxicity by Use of Unreliable Sources and Partisanship}
Having seen the higher toxicity in response to unreliable website submissions, we now examine how the factors previously examined interact with one another to collectively predict toxicity. 

\vspace{2pt}
\noindent
\textbf{Setup.} We fit a linear model to understand how each of the features previously considered (Table~\ref{table:submitter-commeter-info}) predicts the average toxicity comments responding to Reddit submissions. Specifically, we fit our model against the percentage of toxic comments in our 633.59K~unreliable and 7.55M~reliable website submissions. To ensure that our model does not overfit, we run a backward variable selection~\cite{derksen1992backward} based on the Akaike information criterion~\cite{akaike2011akaike} accounting for interaction between our variables. We detail the variables and their found coefficients in Table~\ref{table:toxicity-comments-linear-fit}. 

\begin{table}
   \scriptsize
    \centering
    \selectfont
\setlength{\tabcolsep}{3.5pt}
    \begin{tabular}{lll}
    \toprule
    Variable &  \textbf{Coefficient} & \textbf{Std.}\\    \midrule
    Intercept  &$1.03 \times 10^{-2\scriptscriptstyle{***}}$  & $1.00 \times 10^{-5}$ \\
    Subreddit Toxicity  &$2.80 \times 10^{-3\scriptscriptstyle{***}}$  & $4.20 \times 10^{-5}$ \\
    Subreddit Politicalness  &-$2.00 \times 10^{-4\scriptscriptstyle{*}}$  & $4.66 \times 10^{-5}$ \\
    Commenter Toxicities &$1.02 \times 10^{-2\scriptscriptstyle{***}}$  & $3.74 \times 10^{-5}$ \\
    Commenter Partisanships &-$1.00 \times 10^{-3\scriptscriptstyle{***}}$  & $5.80 \times 10^{-5}$ \\
    Commenter Politcalness &$1.50 \times 10^{-3\scriptscriptstyle{***}}$  & $1.00 \times 10^{-4}$ \\
    Commenter Partisanship:Subreddit Partisanship &$2.00 \times 10^{-4\scriptscriptstyle{***}}$  & $1.00 \times 10^{-5}$ \\
    Commenter Politicalness:Subreddit Politicalness &-$4.00 \times 10^{-4\scriptscriptstyle{***}}$  & $1.00 \times 10^{-5}$ \\
    Moderator of Admin Involved &-$5.00 \times 10^{-5\scriptscriptstyle{***}}$  & $9.37 \times 10^{-5}$ \\
    Is an Unreliable website submission &  $1.30 \times 10^{-3\scriptscriptstyle{***}}$ &  $1.00 \times 10^{-5}$\\
\bottomrule
 \multicolumn{3}{c}{$^\ast p<0.05; \;  ^{**} p<0.01; \; ^{***}p<0.001$}\\
\end{tabular}
  \caption{{Model of the toxicity of the comments in Reddit submissions}. We fit a linear model to model the percentage of toxicity in each of the Reddit threads that contained a reliable domain or an unreliable domain in the submission. We perform backward selection based on the AIC to prevent overfitting. }
   \vspace{-15pt}
   \label{table:toxicity-comments-linear-fit}
\end{table}

\vspace{2pt}
\noindent
\textbf{Results.} As seen in Table~\ref{table:toxicity-comments-linear-fit}, even after accounting for subreddit and user conditions, we see that there is increased toxicity on Reddit in response to an unreliable website submission. Indeed, our model finds this variable to have the fourth largest coefficient ($\beta = 1.30 \times 10^{-3}$) in predicting the overall toxicity of Reddit conversation, behind only overall subreddit toxicity, commenters' propensity for toxicity, and the commenters' politicalness. Our fitted model further finds, as expected from our previous analysis, that moderator involvement is associated with reduced toxicity ($\beta = -5.00 \times 10^{-5}$). This again reinforces that moderator involvement on the Reddit platform is indeed associated with decreased measured toxicity~\cite{trujillo2022make}. As would further be expected, our model determines that subreddit toxicity ($\beta = 2.80 \times 10^{-3}$) and average toxicity of the users that comment ($\beta = 1.02 \times 10^{-2}$) on a given submission is associated with increased toxicity within a given submission's comments. This further shows that the toxicity norms in particular subreddits \emph{do} affect~\cite{rajadesingan2020quick} how users interact.

Our model determines that as subreddits become more political, (\textit{i.e.}, are more aligned along the US political spectrum) overall toxicity decreases. While this result is limited to posts that are centered around news articles, increased politicalness of subreddits in the context of news articles appears to have a slight mitigating effect on toxicity. For example, as previously noted, the r/law subreddit, while not being particularly partisan (-0.19$\sigma$), is over two standard deviations above the mean for politicalness (2.10$\sigma$) and only 0.49\% of the subreddits' comments are considered toxic. We further see that this is the case when examining the interaction between commenter politicalness and subreddit politicalness ($\beta = -4.00 \times 10^{-4}$), and the partisanship of individual commenters ($\beta = -1.00 \times 10^{-3}$). We hypothesize, as found in Rajadesingan et al.~\cite{rajadesingan2020quick} that as subreddits become more aligned to the political spectrum and their users further become aligned to the politicalness of the subreddit or community, stronger community norms are built and overall toxicity decreases. 

However, like Mamakos et~al.~\cite{mamakos2023social}, we find that as the overall partisanship, rather than simply the politicalness of users and subreddits, increases, the toxicity of conversations increases ($\beta = 2.00 \times 10^{-4}$). Finally, again, as in Mamakos et~al.~\cite{mamakos2023social},  we find that as commenters become more political and engage in political discussions ($\beta = 1.50 \times 10^{-3}$), particularly if they engage in both right-leaning and left-leaning discussions and subreddits they tend to have increased toxicity and spread more toxic content on the Reddit platform.

\vspace{2pt}
\noindent
\textbf{Summary.} In this section, after fitting a linear regression model utilizing backward elimination, we find that after accounting for partisanship and other commenter and subreddit-level factors, unreliable website submissions predict increased toxicity on Reddit. Our linear model further identifies that a subreddit's level of political engagement along the US spectrum and toxicity norms also play a role in predicting toxicity.



\subsection{Prediction of Engagement via Toxicity, Use of Unreliable Sources, and Partisanship}
Having shown how the use of unreliable sources predicts increased toxicity on Reddit, we now determine some of the factors that may induce users increased engagement with unreliable websites and their information. Namely, having seen that unreliable sources are associated with more toxicity and more politically right-wing environments compared to reliable sources, are toxicity, politicalness, and partisanship correlated with more engagement with misinformation? 

\vspace{2pt}
\noindent
\textbf{Setup.} To measure user engagement with unreliable and reliable website submissions, we utilize the number of comments that each submission receives.\footnote{We utilize the number of comments rather than the number of upvotes/downvotes because Pushshift often fails to keep up-to-date information about the number of votes for submissions~\cite{baumgartner2020pushshift}.} As before, to properly model the number of comments, we remove comments from Reddit ``auto moderator'' or explicitly ``bot'' labeled accounts. Altogether, we analyze our set of 633.59K unreliable website submissions, our set of 7.55M reliable website submissions, and each of these sets' associated comments. 

To model the number of comments on submissions, we utilize a zero-inflated negative binomial regression~\cite{ridout2001score}. Within our model, each observation data point represents a single submission and its associated number of posted comments. We utilize a zero-inflated negative binomial regression as it appropriately models our set of count data. Unlike a Poisson model, which is often utilized to model count data, negative binomial regressions do not make the strong assumption that the mean of the data is equal to the variance~\cite{morina2022web}. (Some submissions garner thousands of comments while others garner none.) We further utilize the zero-inflated version of this regression given the heavy preponderance of submissions that do not receive any comments. After removing comments from auto moderators and bots, 61.50\% of our reliable website submissions within our dataset did not receive any comments, and 81.67\% of our unreliable website submissions did not receive any comments. A Poisson or normal negative binomial model would be unable to correctly model this behavior. 

We finally note that zero-inflated negative binomial regressions return two sets of coefficients. One set of coefficients, the zero-inflated coefficients, estimated using logistic regression, reports the probability that the given submission would receive zero comments as a function of the covariates. Positive coefficients for these zero-inflated coefficients indicate that increases in the predictor variable make the submissions receiving zero comments more likely. Thus the more negative a coefficient, the more the given covariate correlates with inducing at least 1 comment. The second set of coefficients, the negative binomial coefficients, model the number of comments as a function of the covariates. For these coefficients, positive coefficients indicate that the larger the corresponding covariate, the more comments that submission was likely to have received.  We thus, in our analysis, can understand how different covariates affect the probability that a given submission will receive \textit{any} comments \emph{and} how these same covariates affect the number of comments received. As factors influencing the number of comments, we utilize:
\begin{enumerate}

    \item {the submitter's admin/moderator status}
    \item {the relative age of the account that posted the submission}
    \item {the submitter's partisanship}
    \item {the submitter's politcalness}
    \item {the submitter's account's age}
    \item {the submitter's toxicity}
    \item {the subreddit's partisanship}
    \item {subreddit's politcalness}
    \item {the subreddit's toxicity}
    \item {the average number of comments per submission of the subreddit}
\end{enumerate}
\noindent
We again utilize backward variable selection based on the AIC for selecting variables.\footnote{ We spot-check our results to ensure that the higher the average number of comments in a given subreddit, the more likely a submission is to see comments \emph{and} that this average correlates with more comments on submissions. In other words, we check that submissions in subreddits where users comment more, also see receive comments. As seen in both Tables~\ref{tbl:misinfo-fit} and~\ref{tbl:mainstream-fit}, for both unreliable and reliable website Reddit submissions, as the average number of comments in a subreddit increases, (1) the more likely a submission is to receive comments and (2) the more comments it is likely to receive. Having observed this behavior, we now examine the rest of the covariates within our fits (Tables~\ref{tbl:misinfo-fit} and~\ref{tbl:mainstream-fit}). 
}

\begin{table}[b]
\centering
\scriptsize
\setlength{\tabcolsep}{4.5pt}

\begin{tabular}{lllll}
\toprule

   \multicolumn{5}{c}{\normalsize Number of Comments on Unreliable Website Submissions}          \\
   \toprule
                            & {Zero Inflated}  &  & {Negative Binomial }  \\
                      &       \tiny{negative coefficient = } &  &  \tiny{positive coefficient = } & \\
                      &       \tiny{more likely to get comments} & Std Error &  \tiny{more comments} & Std Error  \\
      \midrule
      Intercept              & 3.30$^{\scriptscriptstyle{***}}$ & 0.14 &0.36$^{\scriptscriptstyle{***}}$  &0.04 \\   
       Submitter Is Moderator & -1.25 $^{\scriptscriptstyle{***}}$& 0.06 & 0.30$^{\scriptscriptstyle{***}}$ &0.04 \\
     Submitter Toxicity &  -19.47$^{\scriptscriptstyle{***}}$ & 1.73 &-0.15 & 0.54 \\
     Submitter Politicalness & -1.14$^{\scriptscriptstyle{***}}$ & 0.24 & -0.49$^{\scriptscriptstyle{***}}$ & 0.09 \\
    Submitter Partisanship       &5.41$^{\scriptscriptstyle{***}}$ &0.34 &     -0.12  &0.12    \\
      Submitter Age & -0.23$^{\scriptscriptstyle{***}}$& 0.01 &0.02$^{\scriptscriptstyle{***}}$ &  0.003 \\
       Subreddit Toxicity  & -0.99$^{\scriptscriptstyle{***}}$ & 0.03& -0.09$^{\scriptscriptstyle{***}}$ & 0.01\\
         Subreddit Politicalness &1.05$^{\scriptscriptstyle{***}}$  &0.03 &0.04$^{\scriptscriptstyle{***}}$ & 0.01 \\
      Subreddit Partisanship    &  0.64$^{\scriptscriptstyle{***}}$ &    0.03 &0.12$^{\scriptscriptstyle{***}}$&   0.01      \\
      |Subreddit Partisanship - Submitter Partisanship| & -0.19$^{\scriptscriptstyle{***}}$ & 0.04& -0.34$^{\scriptscriptstyle{***}}$ & 0.01 \\
      Average \# Subreddit Comments & -2.48$^{\scriptscriptstyle{***}}$ &0.05 & 0.12$^{\scriptscriptstyle{***}}$ &  0.001  \\

\bottomrule
\multicolumn{5}{c}{ $^\ast p<0.05; \;  ^{**} p<0.01; \; ^{***}p<0.001$} \\
\end{tabular}
\caption{Fit of our zero-inflated negative binomial regression on the number of comments on our set of unreliable URL submissions across different subreddits.} 
\label{tbl:misinfo-fit}
\end{table}

\begin{table}[b]
\centering
\scriptsize
\setlength{\tabcolsep}{4.5pt}

\begin{tabular}{lllll}
\toprule

   \multicolumn{5}{c}{\normalsize Number of Comments on Reliable Website Submissions}          \\
   \toprule
                            & {Zero Inflated}  &  & {Negative Binomial }  \\
                      &       \tiny{negative coefficient = } &  &  \tiny{positive coefficient = } & \\
                      &       \tiny{more likely to get comments} & Std Error &  \tiny{more comments} & Std Error  \\
      \midrule
      Intercept              & -3.37$^{\scriptscriptstyle{***}}$& 0.02 & 0.63$^{\scriptscriptstyle{***}}$ & 0.01 \\  
       Submitter Is Moderator &  0.20$^{\scriptscriptstyle{***}}$ & 0.01 & 0.49$^{\scriptscriptstyle{***}}$ & 0.01 \\
     Submitter Toxicity &  -0.06$^{\scriptscriptstyle{***}}$ & 0.003 &-0.02$^{\scriptscriptstyle{***}}$ & 0.002 \\
      
     Submitter Politicalness & 2.43$^{\scriptscriptstyle{***}}$ & 0.02 & 0.22$^{\scriptscriptstyle{***}}$ & 0.006  \\
    Submitter Partisanship       & -0.31$^{\scriptscriptstyle{***}}$ &0.006 &     -0.05$^{\scriptscriptstyle{***}}$  & 0.003    \\
          Submitter Age & -0.15$^{\scriptscriptstyle{***}}$ & 0.004 &0.12$^{\scriptscriptstyle{***}}$ &  0.002 \\
            Subreddit Toxicity  & 0.23$^{\scriptscriptstyle{***}}$ & 0.005 & 0.11$^{\scriptscriptstyle{***}}$ & 0.004\\
     Subreddit Politicalness &0.79$^{\scriptscriptstyle{***}}$  &0.004 &0.19$^{\scriptscriptstyle{***}}$ &  0.002 \\
      Subreddit Partisanship    &  0.51$^{\scriptscriptstyle{***}}$ &    0.004 &0.43$^{\scriptscriptstyle{***}}$ &    0.002      \\
     
      |Subreddit Partisanship - Submitter Partisanship| & 0.46$^{\scriptscriptstyle{***}}$ & 0.006 & 0.08$^{\scriptscriptstyle{***}}$  & 0.003 \\
      Average \# Subreddit Comments & -2.94$^{\scriptscriptstyle{***}}$ &0.01 & 1.60$^{\scriptscriptstyle{***}}$ &0.003  \\

\bottomrule
\multicolumn{5}{c}{$^\ast p<0.05; \;  ^{**} p<0.01; \; ^{***}p<0.001$ }\\
\end{tabular}
\caption{Fit of our zero-inflated negative binomial regression on the number of comments on our set of mainstream URL submissions across different subreddits.} 
\label{tbl:mainstream-fit}
\end{table}

\vspace{2pt}
\noindent
\textbf{Results.} 
We now give an overview and describe some of the implications of our results using our negative binomial regression to predict levels of user engagement based on levels of politicalness, partisanship, toxicity, and the use of unreliable news articles. 
%


\paragraph{Submitter Admin/Moderator Status.} For unreliable website submissions, we find that when a moderator posts the submission they are more likely to get at least one comment compared to a non-moderator account ($\beta = -1.25$).  In contrast, for reliable website submissions, we find that these moderator or admin accounts are less likely to gain at least one comment compared to non-moderator accounts ($\beta = 0.20$). For both unreliable and reliable website submissions, however, we observe that when admin or moderator accounts post, do gain posts, they are more likely to receive more comments than normal accounts. This largely accords with moderators' role on the platform when making announcements in subreddits on which users then comment~\cite{li2022all}.

\paragraph{Submitter Toxicity.}  Examining the submitting users' toxicity, we see somewhat similar behaviors for both reliable and unreliable information submissions. Most notably, as the submitting users become more toxic, for both unreliable and reliable website submissions, they are more likely to provoke at least one comment. However, for unreliable website submissions, we observe that the submitter's toxicity has a much larger effect on the probability of receiving at least one comment($\beta = -19.47$ vs. $\beta = -0.06$). This illustrates that while for unreliable websites, increased toxicity may induce greater initial engagement, this effect is not as strong for reliable websites.  However, again in both cases, we see that while user toxicity often provokes at least one person to react, we see that this toxicity often does not lead to more comments (the coefficient for unreliable websites is not statistically significant). 

\paragraph{Submitter Politicalness.} While we observe that for unreliable websites, the higher a user's politicalness, the more likely to induce at least one comment ($\beta=-1.14$), there is the opposite effect for reliable websites ($\beta=2.43$). This appears to indicate that in the case of reliable website submissions, Reddit users are perhaps being ``turned off'' and are engaging less with highly ideological users compared to less political users~\cite{hetherington2008turned}. However, we also find that the more political a user becomes (if the submission gets comments), the fewer comments unreliable website submissions are likely to receive ($\beta=-0.49$) in contrast to reliable website submissions which receive more comments ($\beta=0.22$). This illustrates that highly politicized users may be more likely to engender a discussion amongst users for reliable website submissions, but are less effective at gathering comments for unreliable website submissions.

\paragraph{Submitter Partisanship.} For unreliable websites, we find that the more right-leaning a user, the less likely the user's post is to attract any user comments. Given the right-leaning nature of most of the subreddits (0.97$\sigma$) in which unreliable domain posts are submitted, this could likely be due to these user's posts being seen as ``normal'' and the posts not receiving many comments ($\beta=5.41$). In contrast, for reliable news  ($\beta=-0.31$), we see that as the submission's submitter becomes more politically right-wing, the more likely their posts are to receive comments. Given reliable website submissions tend to be posted in left-leaning subreddits ($-0.17\sigma$), submissions from more right-leaning users may be seen as more novel resulting in at least one user comment~\cite{kim2021distorting,howard2019ira}. This also supports prior research that has found that out-group animosity may drive online engagement~\cite{rathje2021out}. However, despite right-leaning users being able to attract at least one comment for reliable website submission, we also observe, that as the posting user becomes more right-leaning partisan ideological, the fewer comments their post is likely to receive ($\beta=-0.05$)~\cite{hetherington2008turned}.

\paragraph{Submitter Age.} For both unreliable and reliable websites, we find that older accounts are more likely to provoke at least one comment and that the older the account the more comments that its submission is likely to get. This may indicate that accounts with more history may attract more engagement with their posts.

\paragraph{Subreddit Toxicity.}
Looking at the subreddit toxicity coefficient in predicting whether a submission receives comments, we see a marked difference between reliable website submissions and unreliable website submissions. We see, notably, for misinformation submissions, the more toxic a subreddit is, the more likely the submission is to get comments ($\beta=-0.99$). In contrast, for reliable website submissions, the more toxic the subreddit, the more likely the submission is to not get any comments ($\beta=0.23$). Misinformation websites often post inflammatory articles designed to engender angst in their readership. 

However, we further find, for reliable website submissions, that as subreddit toxicity increases, the more comments submissions are likely to garner. In contrast for unreliable website submissions, the more toxic the subreddit, the fewer comments the submission is likely to garner. This reflects that \emph{when} reliable website submissions get noticed or spark engagement in a toxic community, the more toxic the environment the more users seem to comment and engage with the submissions. In contrast, when articles from unreliable sources are noticed in toxic environments, they do not appear to draw extensive interactions. We thus see that reliable website submissions are more often ignored in toxic subreddits when compared to unreliable websites, and simultaneously that as communities get more toxic, they tend to comment more on reliable information and less on unreliable information submissions.

\paragraph{Subreddit Politicalness.} For both unreliable and reliable website submissions, the more political a subreddit, the fewer users are likely to comment at all ($\beta =1.05$ and $\beta=0.79$. This largely demonstrates again a novelty aspect given that highly political subreddits receive constant news updates. However, for both unreliable ($\beta =0.04$) and reliable submissions($\beta =0.19$), we find that when a submission is commented on, subreddits' politicalnesses increase the likelihood of more comments. This association is again probably largely a result of the fact that work measures users' interaction with reliable and unreliable sources and that subreddits that are more politically engaged on the US political spectrum are more likely to be interested in news~\cite{graber1998politics} and subsequently comment on posts when they gain traction.

\paragraph{Subreddit Partisanship.} We find that for reliable websites, the more politically right-leaning a subreddit, the less likely it is to gain any comments ($\beta=0.51$). Rather, as documented by Wang \textit{et~al.}~\cite{wang2021multi} subreddits like these often ignore more trustworthy sources. We similarly find for unreliable websites, the more politically right-leaning, the less likely these posts are to get any comments ($\beta=0.64$). As before, given the right-leaning nature of most of the subreddits (+0.97$\sigma$) in which unreliable domain posts are submitted, this could likely be due to these users' posts being seen as ``normal''.  In contrast, for both misinformation and reliable website submissions, we find that as the subreddit's right-leaning partisanship goes up,  the more comments given submissions are likely to garner. 

\paragraph{|Subreddit Partisanship - Submitter Partisanship|} For unreliable websites, we find that as the difference between the submitter's partisanship and the subreddit's partisanship increases, the more likely the post is to get at least one comment ($\beta= -0.19$). Various works have found that users not aligned to political norms of a given environment~\cite{rajadesingan2020quick}, provoke engagement from users as they become ``outraged'' by the presented content~\cite{kim2021distorting,gallacher2021online} and can largely be observed here. We note that we do not observe a similar phenomenon for reliable website submissions ($\beta= 0.43$), which may result from the reliable website submission being unable to provoke initial comments. However, for reliable website submissions, we find as the difference between the submitting user's partisanship and the subreddit's partisanship increases, the more comments that that submission is likely to get. This indicates that when the reliable submission manages to gain initiate comments, the farther the submitter's partisanship for the subreddit as a whole, the longer the ensuing conversation. In contrast, for unreliable website submissions, our model finds that as the submitters's partisanship moves further away from the subreddit's own partisanship, after initially accruing an initial comment, it is less likely to gain additional ones.

 \subsection{Summary} 
In this section, we find that submitter toxicity, submitter politicalness, submitter age, subreddit toxicity, and subreddit politicalness all encourage initial interaction with unreliable website submissions. In contrast, submitter toxicity and subreddit toxicity play much more muted roles for reliable news submissions with the subreddit toxicity actually predicting less initial engagement with reliable news sources. This appears to overall suggest a higher degree of initial engagement with unreliable news outlets in political and toxic settings compared to reliable sources.

We further find that moderator involvement, subreddit politicalness, and subreddit partisanship all encourage users to have longer sustained interactions with unreliable information while subreddit toxicity predicts shorter conversations. In contrast for reliable news, we find that subreddit toxicity, subreddit politicalness, and subreddit partisanship all predict increased user engagement (if users initially comment at all). This illustrates that while toxic environments may induce initial engagement with unreliable news, it does not predict sustained interactions,  with the opposite being true of reliable news.


\section{Unreliable Websites and Polarized Toxic Interactions}\label{sec:toxic-polarized-subreddits}
In the previous section, we showed that unreliable website submissions are correlated with increased toxicity and that increased toxicity is also correlated with comments on unreliable website submissions. To understand the user-level dynamics of toxicity in response to unreliable news submissions,  we reconstruct the conversational dyads that exist underneath each Reddit submission. Using the approach outlined in Section~\ref{sec:reddit-methods}, we then determine the partisanship, politicalness, and average toxicity of the users in these conversational dyads, mapping out different types of political interactions. From these averages, we label users as right-leaning (greater than 0.5$\sigma$ partisanship) or left-leaning (less than -0.5$\sigma$ partisanship). Then, looking at each conversational dyad, we determine if each comment is toxic using the Perspective API (as outlined in Section~\ref{sec:reddit-methods}). As an example of such as dyad, in the r/Coronavirus subreddit, a user with a left-leaning bias posted:
\begin{displayquote}
\small
\textit{
Why oh why are people spitting on strangers? And can we get some spit for the evil 80 who own half the planet? No? Ok.}

\end{displayquote}
 to which  another user with a right-leaning bias replied:
\begin{displayquote}
\small
\textit{
Come the fuck on. I don’t care what your opinions are or if it was just a really shitty joke. Don’t wish for people to catch this, that’s an asshole move right there.}
\end{displayquote}
For a comparison of how conversations differ between unreliable website and reliable website comments, we finally separate the set of conversational dyads that appear under unreliable versus reliable website submissions.

\begin{figure}
    \centering
    \begin{subfigure}{.48\textwidth}
       \centering
\begin{minipage}[c]{\textwidth}
   \centering
\begin{tikzpicture}[very thick]
    \node[fill=red!13,draw=black, minimum size=1.2cm, inner sep=0pt] (as) {\footnotesize$1.56\%$};
    \node[fill=red!3,draw=black,minimum size=1.2cm, inner sep=0pt, above=-\pgflinewidth of as] (abh) {\footnotesize$1.33\%$};
    \node[fill=blue!2,draw=black,minimum size=1.2cm, inner sep=0pt, right=-\pgflinewidth of as] (abv) {\footnotesize$1.28\%$};
    \node[fill=red!25,,draw=black, minimum size=1.2cm, inner sep=0pt,  above right=-\pgflinewidth and -\pgflinewidth of as]  {\footnotesize$1.49\%$};
    
    \node[anchor=east,rotate=90,yshift=0.3cm,xshift=0.7cm] at (as.west) {\scriptsize

Right};
    \node[anchor=north,] at (as.south) {\scriptsize
Left};
    \node[anchor=east,rotate=90,yshift=0.3cm,xshift=0.5cm] at (abh.west) {\scriptsize
Left};
    \node[anchor=north] at (abv.south) {\scriptsize
Right};

    \node[xshift=0.5cm, below=0.5cm of as] {\footnotesize Target};
    
    \node[yshift=1.0cm,xshift=-0.35cm,rotate=90,left=0.5cm of as] {\footnotesize Author};

\end{tikzpicture}
\end{minipage}
   \caption{Unreliable Website Submission Dyads}
\end{subfigure}
\begin{subfigure}{.48\textwidth}
\begin{minipage}[c]{\textwidth} 
   \centering
\begin{tikzpicture}[very thick]
    \node[fill=blue!19,draw=black, minimum size=1.2cm, inner sep=0pt] (as) {\footnotesize$1.10\%$};
    \node[fill=blue!9,draw=black,minimum size=1.2cm, inner sep=0pt, above=-\pgflinewidth of as] (abh) {\footnotesize$0.98\%$};
    \node[fill=blue!8,draw=blue!20,draw=black,minimum size=1.2cm, inner sep=0pt, right=-\pgflinewidth of as] (abv) {\footnotesize$1.05\%$};
    \node[fill=blue!23,draw=black, minimum size=1.2cm, inner sep=0pt,  above right=-\pgflinewidth and -\pgflinewidth of as]  {\footnotesize$1.12\%$};
    \node[anchor=east,rotate=90,yshift=0.3cm,xshift=0.7cm] at (as.west) {\scriptsize

Right};
    \node[anchor=north,] at (as.south) {\scriptsize
Left};
    \node[anchor=east,rotate=90,yshift=0.3cm,xshift=0.5cm] at (abh.west) {\scriptsize
Left};
    \node[anchor=north] at (abv.south) {\scriptsize
Right};

    \node[xshift=0.5cm, below=0.5cm of as] {\footnotesize Target};
    
    \node[yshift=1.0cm,xshift=-0.35cm,rotate=90,left=0.5cm of as] {\footnotesize Author};

\end{tikzpicture}
\end{minipage}
 \caption{Reliable Website Submission Dyads}
\centering
 \end{subfigure}
\begin{subfigure}{.48\textwidth}
\begin{minipage}[l]{1.0\textwidth}
\vspace{10pt}
\centering
\begin{tikzpicture}[very thick]
    \node[fill=gray!20,draw=black, minimum size=1.2cm, inner sep=0pt] (as) {\footnotesize
$1.05\%$};
    \node[draw=black,minimum size=1.2cm, inner sep=0pt, above=-\pgflinewidth of as] (abh) {\footnotesize$0.87\%$};
    \node[draw=black,minimum size=1.2cm, inner sep=0pt, right=-\pgflinewidth of as] (abv) {\footnotesize$1.10\%$};
    \node[fill=gray!20,draw=black, minimum size=1.2cm, inner sep=0pt,  above right=-\pgflinewidth and -\pgflinewidth of as]  {\footnotesize$1.03\%$};
    
    \node[anchor=east,rotate=90,yshift=0.3cm,xshift=0.7cm] at (as.west) {\scriptsize

Right};
    \node[anchor=north,] at (as.south) {\scriptsize
Left};
    \node[anchor=east,rotate=90,yshift=0.3cm,xshift=0.5cm] at (abh.west) {\scriptsize
Left};
    \node[anchor=north] at (abv.south) {\scriptsize
Right};

    \node[xshift=0.5cm, below=0.5cm of as] {\footnotesize Target};
    
    \node[yshift=1.0cm,xshift=-0.35cm,rotate=90,left=0.5cm of as] {\footnotesize Author};

\end{tikzpicture}
\end{minipage}
 \caption{All Submission Dyads}
 \end{subfigure}
   \caption{Percentage of interactions that are toxic in all, unreliable, reliable website submissions for Right and left-leaning authors against Right and left-leaning targets.\label{fig:increase-partisanship-dyads} }
\end{figure}   

\subsection{Interactions within Unreliable and Reliable Information Ecoysystems\label{sec:environment}}

We observe (as expected) that many users primarily interact with users of the same partisanship~\cite{stromer2003diversity}: 71.80\% of interactions were between users that share the same partisanship-lean. For unreliable news submissions, this rises to 83\%, and for reliable website submissions, it drops to 66\%. We similarly find that 72.08\% of toxic interactions (where a user responded to another user with a toxic reply) were between users who shared the same partisanship leaning among all dyads, 80.63\% for unreliable website submissions, and 63.34\% for reliable website submissions. This is likely because, as previously found, unreliable domains are largely posted in somewhat more insular subreddits (average partisanship = 0.97$\sigma$; Table~\ref{table:submitter-commeter-info}) and in communities with higher degrees of toxicity (2.01\%; Table~\ref{table:submitter-commeter-info}). 

Despite users largely interacting with users of the same partisanship, we find some increased rates of affective polarization between users of different partisanships. As seen in Figure~\ref{fig:increase-partisanship-dyads}, we observe increased toxicity between users of different partisanships for our set of website submissions, with this difference most marked for unreliable website submissions. Indeed calculating the odds ratios between the percentages of inter-partisanship toxicity against those of intra-partisanship toxicity, we get values of 0.99 across all dyads, 1.19 for unreliable domain dyads, and 1.08 for reliable domain dyads. We thus observe a slight increase in inter-partisanship toxicity between users who comment under submissions with attached domain hyperlinks. Further, calculating the odds ratio between the rates of toxicity between unreliable websites and reliable website conversational dyads, we get values of 1.38 for inter-partisanship toxicity and 1.26 for intra-partisanship toxicity. We thus observe that amongst our set of conversations, there is an even heightened rate of affective polarization for unreliable news interactions compared to reliable news interactions. 

\subsection{Modeling Toxic Interactions Between Users}

To concretely show that users of different political stripes are more likely to reply in a toxic manner to each other in conversations under unreliable domain submissions, we fit our network data of toxic interactions into an exponential random graph model. An Exponential Random Graph Model (ERGM) is a form of modeling that predicts connections (\textit{e.g.}, toxic interactions) between nodes (users) in a given network~\cite{hunter2008ergm}. ERGM models assume that connections are determined by a random variable $p^*$ that is dependent on input variables. As in Chen \textit{et al.}~\cite{chen2022misleading} and Peng \textit{et al.}~\cite{peng2016follower}, we utilize this modeling as it does not assume that its data input is independent; given that, we want to model the interactions of polarization, toxicity, this relaxed restriction is key (we have already seen that they are largely not independent)~\cite{van2019introduction,hunter2008ergm}.

\vspace{2pt}
\noindent
\textbf{Setup.}  Utilizing our ERGM, we predict the probability of toxic interactions between two users within misinformation submissions as a function of:
\begin{enumerate}

    \item {the users' percentage of toxic comments}
    \item {the users' partisanship}
    \item h{the difference in the author and target's political polarization}
    \item {the users' politicalness}
    \item {the age of the two users}
    \item {the reciprocity between the two users (\textit{i.e.}, if both users had a toxic comment aimed at each other)}
    \item {the number of comments that the two users have in subreddits in which they both post comments}
\end{enumerate}
\noindent
We include the number of comments that the users have made in shared subreddits to account for the fact that users with more overlap in user activity (\textit{i.e.}, frequent the same subreddits) are more likely to interact with one another. When fitting our models we again utilize backward selection and minimize the AIC to determine the variables used in our final models.

\begin{table}[t]
\centering
\scriptsize
\begin{tabular}{lll}
\toprule
  \textbf{Unreliable  Domain Interactions }  &  \textbf{Coeff.}  & \textbf{Std.}  \\ \midrule
      Intercept               &8.65*** & 0.05 \\               
      User Partisanship Differences     & -0.20***& 0.04 \\
       User Toxicity & 5.88*** & 0.46\\
       Shared Subreddits Comments &  0.004*& 0.001\\
       Reciprocity &  4.79*** & 0.18\\
\bottomrule
\multicolumn{3}{c}{ $^\ast p<0.05; \;  ^{**} p<0.01; \; ^{***}p<0.001$ }\\
\end{tabular}
\begin{tabular}{lll}
\toprule
 \textbf{Reliable Domain Interactions}  &  \textbf{Coeff.} & \textbf{Std.}  \\ \midrule
      Intercept               &8.73***& 0.05 \\               
      User Partisanship Differences     & -0.29***  & 0.04\\
       User Toxicity &  6.48*** & 0.74\\
       Shared Subreddits Comments &  0.001*** & 0.0004\\
       Reciprocity & 3.97*** & 0.27\\
\bottomrule
\multicolumn{3}{c}{ $^\ast p<0.05; \;  ^{**} p<0.01; \; ^{***}p<0.001$ }\\
\end{tabular}
\caption{{Toxic Unreliable and Reliable Website Submission Interactions}. As confirmed in our ERGM, differences in the political orientation of users are predictive of increased incivility and toxicity, with users of differing political orientations more likely to engage in toxic interactions within misinformation submissions than on mainstream submissions. Similarly, the higher each user's toxicity norm, the more they are likely to target other users with toxic comments. }
\label{tbl:ergm}
\vspace{-15pt}
\end{table}

\vspace{2pt}
\noindent
\textbf{Results.} We find that account age, partisanship, and the politicanlness of a given user do not have significant effects on the likelihood of toxic interactions (removed from fit after minimizing the AIC). This indicates that just because a user is highly partisan or political it does not necessarily mean that they are likely to engage in toxicity. For all domain interactions, as seen in Table~\ref{tbl:ergm}, we find that (1) that the more toxic a user, the more likely they are to engage in toxic interactions, and (2) that users are more likely to respond in a toxic manner to users who engage with them in a toxic manner (reciprocity). Indeed we find that in unreliable website submissions, users are more likely to reply in a toxic manner to another user if that user has already corresponded with them in a toxic manner ($\beta=4.79$ vs. $\beta= 3.97$). However, most importantly, we find that while most toxic interactions occur among users that are politically similar to each other, compared to reliable domain interactions, users discussing unreliable website submissions are \emph{more} likely to send toxic comments to users of different political ideologies than users under mainstream submissions ($\beta=-0.20$ vs. $\beta= -0.29$).


\vspace{2pt}
\noindent
\textbf{Summary.} In this section, we showed that unreliable website submissions not only promote higher levels of toxicity in their comments but are also correlated with increased inter-partisanship toxicity compared to reliable website submissions. Fitting an ERGM to our toxic conversational dyads posted in response to misinformation stories, we show that political differences, along with reciprocity and each user's toxicity, drive more toxic interactions.

\section{Limitations}
In this work, we used a quantitative, large-scale approach to understand the role of misinformation in toxic interactions online. We outline the limitations of our approach in this section.

\vspace{3pt}
\noindent
\textbf{Unreliable Information.}
One of the limitations of our approach is our use of hyperlinks to determine the presence of unreliable/factually inaccurate information. As we examined much of Reddit's 2.2~billion comments, we were unable to take a comment-by-comment-based approach to understand the levels of unreliable news. As a result, our approach inevitably missed some subtleties of unreliable information across subreddits. However, as found in several past works~\cite{hanley2022no,hounsel2020identifying,sehgal2021mutual,waissbluth2022domain}, examining unreliable information from a domain-based perspective enables researchers to track readily identifiable and questionable information across different platforms and is a reliable way of understanding the presence of unreliable information in large communities or websites (\textit{e.g.}, subreddits). Our approach thus relies on the presence of largely US-based domains on given subreddits and largely only measures English unreliable information and partisanship. As a result, we cannot simply apply our results to non-English subreddits and non-US-oriented environments.  However, we note, that while our work centers on US-based political environments, as found in prior works, highly political environments across different cultures often utilize unreliable information and often share many of the same characteristics as US ones~\cite{imhoff2022conspiracy, hanley2022no}. We leave the full investigation of this phenomenon on Reddit to future work.

\vspace{3pt} 
\noindent
\textbf{Measuring Toxicity.}
Another limitation of our approach, given our use of the Perspective API to estimate toxicity, is that it is limited to relatively active users and subreddits. We are only able to develop, in line with past works, toxicity norms and political estimations for subreddits that have at least 100~comments. As such, our results are skewed to more active subreddits and users. At the same time, these subreddits and users make up a large percentage of users' experiences on Reddit. 

\vspace{3pt}
\noindent
\textbf{Confounds, Correlation, and Causation.}
We lastly acknowledge that while we account for many user-level and subreddit-level features, there may be other hidden confounders. For example, while we attempted to remove automated accounts from much of our analysis by removing accounts that were labeled as ``bot'' accounts, due to the rapid rise of AI, within Reddit as a whole there could still be automated accounts. We note that we conducted this analysis for data in 2020 and 2022, before the release of ChatGPT however. We further emphasize that while we work to account for confounders, the results we present describe the correlation between misinformation, political polarization, and toxicity; we cannot ascribe causation. However, our results do align with a large literature of similar results~\cite{barbera2014social,barbera2015tweeting,bail2018exposure} some of which have found causal results.

\section{Discussion}
In this work, we examined the relationship between unreliable information, political partisanship, user engagement, and toxicity across and within both political and non-political subreddits. Using previously published lists of unreliable and reliable websites, we find that on Reddit, we find that comments posted in response to submissions with hyperlinks to unreliable news websites have 25\% more toxic comments overall (an average of 32\% more within individual subreddits). Utilizing a zero-inflated negative binomial model to model engagement with unreliable versus reliable information sources, we observe that subreddit toxicity is a major predictor of whether unreliable domain submissions receive comments. This contrasts with reliable domain submissions, where toxicity plays a more muted role, and the more toxic the subreddit, the more likely that reliable submissions are to not get any comments. Finally, examining how partisanship affects the increase in toxicity in response to unreliable information, we find, confirming with an Exponential Random Graph Model (ERGM), that articles from unreliable news outlets correlate with increased toxicity among users of different political leanings (\textit{i.e.}, affective polarization).

\subsection{Unreliable Information's Correlation with Toxicity}
Our work shows that while unreliable websites have much less of a presence on Reddit compared to reliable websites (633.6K posts/601 submissions per domain vs 7.55M posts/2010.4 submissions per domain), unreliable news websites play a large role on the platform. As documented by others, often millions of comments discuss and spread false information~\cite{setty2020truth}. In addition to misleading users, unreliable information's effect on the discourse on these subreddits can often be pernicious with articles from websites known to promote unreliable news increasing inter-political strife. Indeed as was seen in Table~\ref{table:submitter-commeter-info} and was found in our unreliable domain submission dyads, unreliable domain submissions are associated with increased toxicity, particularly among users of different partisanship alignments. This largely accords with the work of Dicicco {et~al.}~\cite{dicicco2020toxicity} that showed that users who comment on YouTube videos promoting COVID-19 conspiracy theories often utilize toxic and vulgar language. Our paper results bolster this work, showing that increased unreliable domains correlate with increased incivility on Reddit. This largely goes to promote and affirm the view that unreliable news/misinformation does have a relationship~\cite{mosleh2024misinformation,dicicco2020toxicity} with user toxicity and is not uncorrelated with toxicity~\cite{cinelli2021dynamics,quattrociocchi2022reliability}.

In our conversational dyads, we further find that across much of Reddit, unreliable websites are correlated with more insular and politically one-sided conversations, while reliable domains are correlated with increased discussions between users of different political ideologies (with both increasing inter-political toxicity). Community norms for particular environments appear to affect how users engage with different materials. As found with our zero-inflated negative binomial model, subreddit toxicity norms are also predictive of user engagement with unreliable news articles. Unreliable and factually inaccurate, is found within toxic environments. The more toxic/uncivil a given environment, the more likely at least one person is to engage with misinformation or unreliable sources. However, simultaneously, in more toxic environments, where these posts most commonly appear, these same posts are less likely to gain extensive engagement and a large number of comments. This appears to reflect that unreliable news websites often utilize ``clickbait'' titles that induce readers to initially comment, but then cause the reader to not often thoroughly engage with material~\cite{chen2015misleading,potthast2016clickbait}. In contrast, in less toxic environments where these posts more rarely appear, if they do gain traction (\textit{e.g.}, at least one comment), they are more likely to gain more comments. 

\subsection{Implications of the Reddit Platform}
Our work indicates that unreliable domains correlate with increased overall toxicity of conversations on Reddit, particularly between users of different partisanships. We note that this increased rancor persists despite individual subreddits moderating unreliable domain submissions more heavily compared to reliable domain submissions. Given the lower prevalence of unreliable sources throughout Reddit compared to reliable sources and the decreased toxicity of conversations with moderator involvement, a potential solution to decrease toxicity may be for Reddit admins (who are not already doing so), to engage more thoroughly or to flag submissions that contain hyperlinks to known unreliable and specious websites. However, as argued by Bozrath et~al.~\cite{bozarth2023wisdom}, different approaches for moderating this content in different subreddits however will be necessary. Some larger subreddits already take a machine-learning approach to remove misinformation~\cite{jhaver2019human} while others take a manual approach that relies on crowd wisdom or individual moderator involvement~\cite{seering2018social,hwang2021people,jones2019r}. However, given that Reddit removed links to Russian state-based propaganda in the wake of the Russo-Ukrainian War~\cite{Spangler2022} and has previously taken steps to remove highly toxic material and subreddits~\cite{Spangler2020}, we recommend that Reddit itself also take more proactive steps to alert users to unreliable information and to identify new websites and known websites that promote unreliable information and flag, label, or remove them from their platform. Further as again found by Bozrath et~al.~\cite{bozarth2023wisdom} moderating one type of misinformation or unreliable source can be similar to moderating other types, allowing Reddit to take a generalized approach to alert subreddits to the presence of unreliable news and propaganda.

\vspace{2pt}\noindent
\textbf{Political Echo-Chambers, Politics Discussions, and Reliable News on Reddit.}
Similar to past work, we find that most toxic interactions take place among users of the same political orientation~\cite{efstratiou2022non}. Reddit specifically creates communities for like-minded people and as a result, most interactions (both toxic and non-toxic interactions) on the platform are between people of the same political orientation. However, most interestingly, in the comments of submissions with hyperlinks to reliable news sources, the rate of inter-partisan interactions slightly increases compared to interaction across Reddit. This is in contrast to unreliable domain submissions where the rates of interaction between users and different partisanship decreases.  We argue, that if Reddit, as a whole, desires to lower levels of political incivility and toxicity on its platform, taking a more proactive approach to policing questionable sources could help alleviate these issues. As found by Gallacher {et~al.}~\cite{gallacher2021online}, toxic online interactions between political groups often lead to offline real-world political violence. Given that unreliable news appears to be correlated with and reinforces toxic interactions between different political groups, this highlights the need to research its effects and curtail its spread.

\vspace{2pt}\noindent
\textbf{Sub-Standards/Community Norms.}
We have found throughout this work that subreddits interact with reliable and unreliable sources differently. For example, while more toxic subreddits are more likely to interact with unreliable information sources, the more toxic a subreddit, the more likely the reliable submissions are to not get any comments. We thus find often complex relationships between different types of subreddits and their interactions with different types of posts. There is no one-size-fits-all approach to understanding user engagement and toxicity on Reddit~\cite{seering2022metaphors,zhang2021understanding}.  We thus argue that a subreddit/community-based approach that takes into account the community norms of the community must be taken when trying to understand the information flows within it~\cite{fiesler2018reddit}. Similarly, in attempting to prevent engagement with unreliable news on particular subreddits, understanding their toxicity norms, their political ideology, and who is posting the article within the subreddit is key~\cite{zhang2021understanding}. For example, as found by Zhan et al~\cite{zhang2021understanding}, different communities responded and engaged with COVID-19 misinformation in widely divergent manners. We thus argue that approaches that attempt to understand how users engage with unreliable information (particularly on Reddit), \emph{must} take into account the particular nuances of that community.

\section{Conclusion}
Unreliable information persists across many different types of subreddits. Its spread furthermore seems to be affected by the type of community it is posted in. Unreliable and factually incorrect appears to be more likely to gain traction when it is posted in more toxic/uncivil environments. Furthermore, the communities with large amounts of unreliable news appear to be more politically insular with more of their interactions occurring between users of similar political orientations. As users become more politically dissimilar when commenting under unreliable information, as found with our ERGM, they are more likely to be toxic/uncivil to one another compared to users who comment under reliable information.  Our work, one of the first to examine the relationship between unreliable news, toxicity, and political ideology at scale, illustrates the need to fully understand the full effect of unreliable information. Not only does unreliable news mislead people but it also can magnify political differences and lead to more toxic online environments. 


\bibliographystyle{ACM-Reference-Format}
\bibliography{sample-base}

\appendix
\section{Embeddings Hyperparameter Optimization\label{sec:hyperparamters}}
\begin{table}[!h]
    \scriptsize
    \centering
    \begin{tabularx}{0.65\columnwidth}{l|l}
    \toprule
    Variable &  Values Considered\\    \midrule
    Embedding Size & 100, 150, 200 \\ 
    Number of negative examples  & 30, 35, 40, 45 \\
    Down-Sampling threshold; & 0.0025 0.005, 0.0075, 0.01 \\
    The starting learning rate & 0.15, 0.18, 0.21 \\
\bottomrule
    \end{tabularx}
  \caption{We optimize our community and user embeddings. }
   \vspace{-15pt}
   \label{table:emebeddings-otpimization}
\end{table}

\end{document}